\documentclass[twocolumn,english,aip,rsi,reprint]{revtex4-1}
\usepackage[T1]{fontenc}
\usepackage[latin9]{inputenc}
\usepackage{amsmath}
\usepackage{graphicx}

\makeatletter
 \@ifundefined{textcolor}{}
 {%
   \definecolor{BLACK}{gray}{0}
   \definecolor{WHITE}{gray}{1}
   \definecolor{RED}{rgb}{1,0,0}
   \definecolor{GREEN}{rgb}{0,1,0}
   \definecolor{BLUE}{rgb}{0,0,1}
   \definecolor{CYAN}{cmyk}{1,0,0,0}
   \definecolor{MAGENTA}{cmyk}{0,1,0,0}
   \definecolor{YELLOW}{cmyk}{0,0,1,0}
 }

\makeatother

\usepackage{babel}
\begin{document}

\title{Arbitrary Waveform Generator for Quantum Information Processing with
Trapped Ions}

\author{R. Bowler}

\email{ryan.bowler@nist.gov}

\address{National Institute of Standards and Technology, Boulder, CO 80305,
USA}

\author{U. Warring}

\address{National Institute of Standards and Technology, Boulder, CO 80305,
USA}

\author{J. W. Britton}

\address{National Institute of Standards and Technology, Boulder, CO 80305,
USA}

\author{B. C. Sawyer}

\address{National Institute of Standards and Technology, Boulder, CO 80305,
USA}

\author{J. Amini}

\altaffiliation{Work was performed while this author was at the National Institute of Standards and Technology.}

\address{Georgia Tech Research Institute, Atlanta, GA 30332, USA}
\begin{abstract}
Atomic ions confined in multi-electrode traps have been proposed as
a basis for scalable quantum information processing. This scheme involves
transporting ions between spatially distinct locations by use of time-varying
electric potentials combined with laser or microwave pulses for quantum
logic in specific locations. We report the development of a fast multi-channel
arbitrary waveform generator for applying the time-varying electric
potentials used for transport and for shaping quantum logic pulses.
The generator is based on a field-programmable gate array controlled
ensemble of 16-bit digital-to-analog converters with an update frequency
of 50~MHz and an output range of $\pm$10~V. The update rate of
the waveform generator is much faster than relevant motional frequencies
of the confined ions in our experiments, allowing diabatic control
of the ion motion. Numerous pre-loaded sets of time-varying voltages
can be selected with 40~ns latency conditioned on real-time signals.
Here we describe the device and demonstrate some of its uses in ion-based
quantum information experiments, including speed-up of ion transport
and the shaping of laser and microwave pulses.
\end{abstract}
\maketitle

\section{Motivation}

The building blocks for a quantum information processor involve qubits
(two-level quantum systems), a universal set of operations for quantum
logic, and transport of information about the processor. One realization
of a quantum information processor is to use internal states of ions
as qubits, where the ions are confined in a Paul trap via a combination
of static and RF electric fields that forms a three-dimensional harmonic
oscillator potential \citep{95Cirac,98Wineland}. By designing a trap
with multiple electrodes, arrays of trapping locations can be created
between which ions can be transported. Transport is achieved via the
application of time-varying electric potentials to the electrodes.
Ions can then be transported between zones dedicated to particular
operations such as state detection or quantum logic \citep{00Cirac,02Kielpinski}.
The latter is implemented by use of a sequence of laser or microwave
pulses for excitations between the two levels of internal electronic
states of the qubit ion.

Demonstrations of this multi-electrode trap array scheme using a suitable
pair of $^{2}$S$_{1/2}$ hyperfine states of $^{9}$Be$^{+}$ as
qubits at the National Institute of Standards and Technology (NIST)
involved ion transport on a long time scale \citep{10Hanneke,12Gaebler}.
While each quantum logic operation typically took $\sim10\,\mu$s,
ion transport required hundreds of microseconds to remain in the adiabatic
regime and thus suppressed unwanted excitations of ion oscillatory
motion. A demonstration of an error correction algorithm, which is
conditioned on the detection of errors, has also been carried out
in this multi-electrode scheme \citep{04Chiaverini}. For larger scale
quantum information processing, development of larger trap architectures
is necessary; some examples of the work towards larger multi-electrode
trap designs can be seen in Refs.\citep{08Huber,10Amini,11Blakestad,12Allcock,12Doret}.
However, the time scale discrepancy between quantum logic and ion
transport poses a bottleneck for rapid quantum information processing.
In addition, motional excitation due to electric field noise from
electrode surfaces \citep{00Turchette}, the need for laser cooling
to initialize motion close to the quantum ground state (via Doppler
and Raman sideband cooling \citep{98Wineland}), and the long duration
of adiabatic transport poses a bottleneck if a single transport and
cooling sequence exceeds a few milliseconds.

\begin{figure}[t]
\includegraphics{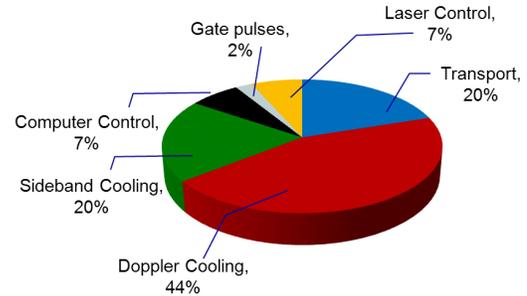}

\caption{The distribution of durations between different tasks within a typical
quantum logic process in Ref.\citep{12Gaebler}. The sum of quantum
logic pulses amounts to a 100~$\mu$s duration, which is only 2~\%
of the total duration. Computer control includes preparing electronics
for the experimental sequence (including the preparation of DAC output
patterns for transport), laser control sets frequencies and beam powers
for quantum logic, and the largest portions are due to transport and
the initialization of the motional states through Doppler and Raman
sideband cooling.}

\label{figure:PieChart}
\end{figure}

Time scales in Refs.\citep{10Hanneke,12Gaebler} were primarily dominated
by slow ion transport and cooling of ion motion, followed by computer
control (including the preparation of electronics for transport) as
another dominant temporal overhead, diagrammed in Fig.~\ref{figure:PieChart}.
Previous experiments at NIST used commercially available digital-to-analog
converters (DACs) with update rates of 500~kHz for ion transport,
but update rates lower than the frequencies of ion motion in the confining
potential well are unsuitable for precise control of the potential
on time scales similar to or faster than the ions' motional period.
In addition, aliased harmonics of the 500~kHz update rate could cause
significant heating of the ions' motion if not heavily filtered \citep{11Blakestad}.
Large-scale error correction algorithms may require the ability to
quickly alter ion transport conditioned on the detection of errors.
Hence, our electronic needs for this quantum information processor
scheme include fast DAC update rates compared to confinement frequencies,
which can be used to reduce the duration of ion transport with low
motional excitations on a quantum logic time scale, the reduction
of computer control duration, the ability to control many electrodes
at once, and low latency in accessing DAC output patterns (waveforms)
which can be conditionally selected.

We have addressed our current technical needs by creating a fast multi-channel
arbitrary waveform generator (AWG) with an update rate of 50~MHz,
well beyond currently relevant trapping frequencies (typically 1 to
3~MHz). The design is easily scalable to control a large number of
electrodes. We report on the design details and operation of the AWG
and describe a few of its applications with trapped ions. The AWG
is based on a field-programmable gate array (FPGA), which controls
an ensemble of DACs that provide arbitrary analog outputs in a $\pm$10~V
range. The device is capable of simultaneously storing multiple ``branches'',
different sets of waveforms that can be selected by use of digital
logic inputs on the device with a 40~ns latency. We conclude with
descriptions of experiments that demonstrate the use of AWGs in fast
transport, multiple waveform branching for transport, and pulse shaping
applications for ion-based quantum information processing.

\section{Design}

\subsection{Hardware}

\begin{figure}[t]
\includegraphics{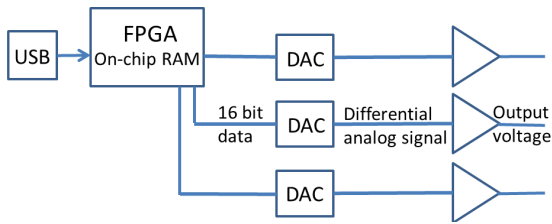}\caption{A flow diagram overview of the AWG. The host computer transmits waveform
data via USB to the FPGA which is stored in on-chip RAM. The FPGA
transmits a 16-bit digital signal derived from its RAM to each DAC,
whose output is then amplified.}

\label{figure:FlowDiagram}
\end{figure}
The AWG uses the Xilinx Spartan-3E XC3S500E PQ208 \citep{TradeDeclaration}
model FPGA to control three 16-bit Analog Devices AD9726 \citep{TradeDeclaration}
DACs. The FPGA logic was coded in VHDL using the Xilinx ISE Design
Suite \citep{TradeDeclaration}. The AD9726 DAC was chosen because
of its high speed, its linearity, and its use of low voltage differential
signal (LVDS) which is resistant to pick-up noise. A host computer
provides waveform specifications that are stored in three segments
of independent static random access memory (RAM) in the FPGA for each
connected DAC channel (provided by a total of 360 kilobits of on-chip
RAM). The FTDI FT245RL USB-to-parallel-FIFO \citep{TradeDeclaration}
provides an interface to connect a host computer to the FPGA memory,
across which waveform data is uploaded to RAM. An overview for the
operation of the AWG is shown in Fig.~\ref{figure:FlowDiagram}.

The AWG has a USB connection to the host computer and a 50~MHz crystal
oscillator to provide a clock signal. There are two types of AWGs,
a ``master'' and a ``slave'', with a small number of hardware
differences. Slaves are connected to the master through board-to-board
interconnects, through which the waveform data and the clock signal
are routed to be processed by their respective FPGAs. One master and
two slaves can be connected to provide up to nine synchronous channels
on a single USB connection. Separate masters are referenced to different
oscillators and hence have timing offsets of at most 20~ns, though
this is so far negligible in our experiments. Each AWG master is identified
over the USB connection by a unique serial number programmed into
the FT245RL, so in principle any number of channels can be controlled
from a single computer (limited by the computer's USB architecture).

Groups of each channel's RAM can be configured into eight independent
waveform branches of user-defined sizes externally selected with three
TTL user inputs. The TTL inputs switch waveforms with 40~ns (two
clock cycles) latency and allow the user to switch between waveforms
in real-time with the rest of the experiment in response to events.
There is an additional input TTL trigger on each master for signaling
the start of a waveform on all connected channels; the rising edge
triggers the start of reading waveform data from the selected branch
on all channels. By tying more than one master to a single TTL, many
masters (and their respective slaves) can be run synchronously. Each
waveform branch may further be broken up into a sequence of waveform
``steps'', such that in a single branch a set of TTL triggers will
run a sequence of waveforms.

\begin{figure}[t]
\includegraphics{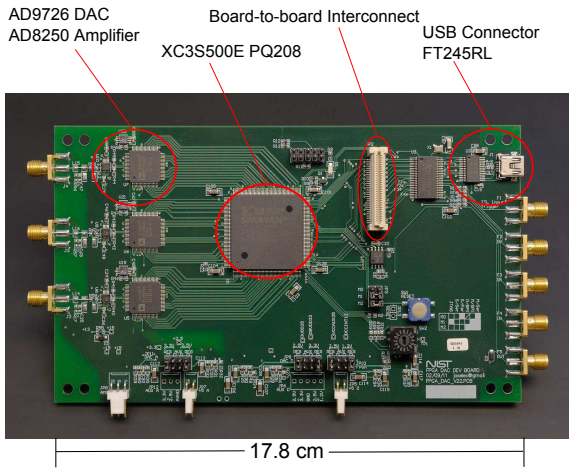}

\caption{A master board. The key components described in the text are highlighted.
The host computer communicates with the master board via USB. The
FT245RL handles the interface with the XC3S500E PQ208 FPGA on up to
three boards (a master and two slaves) on a shared communication channel
through a board-to-board interconnect. The FPGA drives three AD9726
DACs whose output is fed to AD8250 amplifiers, which creates the final
three output potentials.}

\label{figure:MasterBoard}
\end{figure}

The DACs are clocked from a 50~MHz signal generated by the FPGA.
Each DAC's output is referenced to a precision 1.25~V source (Linear
Technology LTC6652 \citep{TradeDeclaration}), specified to have 10~ppm/$^{\circ}\text{{C}}$
and 2.1~ppm noise, for stable and consistent potentials out of every
channel. The differential signal outputs of each DAC are then sent
to high-speed amplifiers (Analog Devices AD8250 \citep{TradeDeclaration})
to provide the final output potentials; the DAC and amplifier system
is configured for a $\pm$10~V output with 16-bit linear resolution,
stable down to roughly 30~$\mu$V and with $\sim1\,\Omega$ output
impedence. A 50~$\Omega$ resistor can be added in series with the
output to prevent ringing in long cables. After the resistor we observe
a white noise power spectral density of $\sim$120~nV/$\sqrt{{\rm Hz}}$
with an external 50~$\Omega$ load. The amplifier's output has a
measured slew rate of 40~V/$\mu$s; although this limits the voltage
swing at each clock update, it is sufficient for our experimental
implementations. A photograph of a master board is shown in Fig.~\ref{figure:MasterBoard}
with the discussed components highlighted.

\subsection{Software}

The host computer can specify the waveforms in one of four modes as
follows: (1) Output specified by a potential during each 20~ns clock
cycle update, (2) Output given by potentials for specified durations,
(3) Output given in terms of quadratic polynomials specified by a
duration and coefficients, and (4) Output specified in terms of cubic
polynomials (of primary interest, cubic spline interpolation for $1^{st}$
and $2^{nd}$ derivative continuity in the waveform). Options 2, 3,
and 4 can provide increasing levels of data compression. This is often
necessary to overcome limited amounts of RAM and reduce waveform upload
durations from the host computer. When operating in the $4{}^{th}$
mode, up to 614 polynomials can be stored in each channel\textquoteright{}s
RAM. The interval between a TTL trigger and when a waveform begins
to run depends on the time required to read the initial data from
memory (longest is the $4{}^{th}$ mode where 240~ns, or twelve clock
cycles, are required before the waveform begins), but after this initial
delay the memory is read continuously while the waveform runs so there
is no slow-down in run time.

Our in-house data acquisition system provides control of parameters
in our experiments: the timing and sequences of most quantum logic
pulses via a separate FPGA timing system, the waveform upload to the
AWGs (including specifications for the channel and branch configuration
for each waveform), and any other dynamic experimental parameters
(see Ref.\citep{06Langer} for details). This system controls experimental
TTL signals including the run trigger and branch switching logic for
the AWG, along with switching and timing TTL signals for various quantum
logic pulses. The FT245RL has a support library of C++ commands provided
by the manufacturer, which is used in the data acquisition system
for communication between the host computer and the AWGs.

\section{Experimental Applications}

\subsection{Fast Ion Transport}

\begin{figure}[t]
\includegraphics{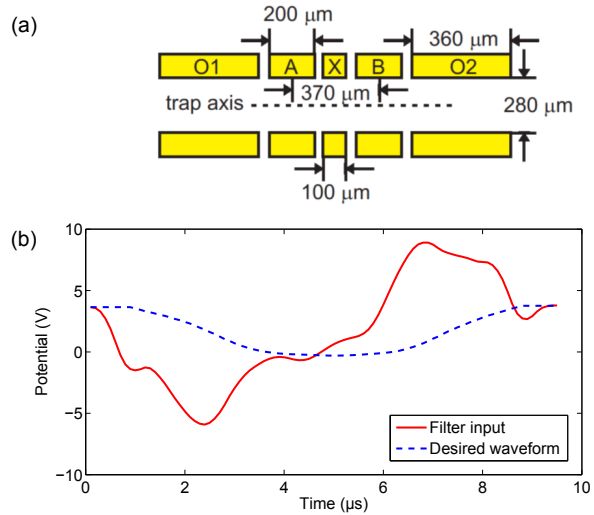}

\caption{(a) Schematic of the ion trap electrode structure showing the two
diagonally opposite segmented DC electrodes (not to scale). Outer
electrodes O1 and O2 are used to fine tune the potential while ions
move over electrodes~A, X, and B during the experiments in Ref.\citep{12Bowler}.
(b) Example filter compensation for electrode~X during transport
of an ion from over electrode~A to over electrode~B. The desired
waveform is generated from simulation and the potentials generated
by the AWG must account for filter distortion. The waveform from the
AWG must overshoot in time and voltage to achieve the desired 8~$\mu$s
transport potentials.}

\label{figure:FilterCompensation}
\end{figure}
For rapid quantum information processing in a multi-electrode trap
architecture, ion transport should occur on a time scale comparable
to quantum logic (e.g. $\sim10\,\mu$s for ion based systems) or faster.
The high speed of the AWG relative to confinement frequencies in a
linear Paul trap enables variation of electrode potentials for ion
transport within a fraction of the ion oscillation period. We employed
the AWG for fast, diabatic transport of ions \citep{12Bowler}, where
the transport waveforms accelerated and decelerated the ions within
the ion oscillation period and observed a low final motional excitation
in a single transport. In our experiments, the AWG produced the time-varying
electric potentials in a multi-electrode linear Paul trap (diagrammed
in Fig.~\ref{figure:FilterCompensation}a) to transport one and two
$^{9}$Be$^{+}$ ions in 8~$\mu$s across 370~$\mu$m at a constant
velocity, with the ions confined in a $\omega_{z}/2\pi\sim2$~MHz
potential well. We also separated two $^{9}$Be$^{+}$ originally
held in a single confining well in 55~$\mu$s into two separate trap
zones. These experiments demonstrated that transport and separation
over relevant distances on quantum logic time scales can be performed
with suppressed excitation in the trapped ions' final motional state.

\begin{figure}[t]
\includegraphics{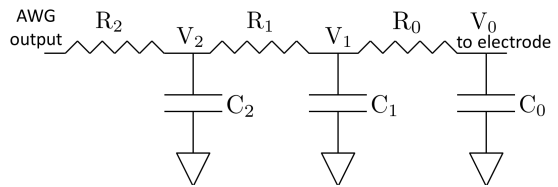}

\caption{A diagram of our low-pass RC filter system connected to each electrode.
The indices of each passive component correspond to the filter layer,
and the voltages correspond to those governed by Eq.~\ref{eq:Filter}
with V$_{0}$ the final stage applied to the electrode. The values
are \{R$_{2}$, R$_{1}$, R$_{0}$\} = \{820, 820, 240\}~$\Omega$
and \{C$_{2}$, C$_{1}$, C$_{0}$\} = \{1, 1, 0.82\}~nF.}

\label{figure:trapfilters}
\end{figure}

A complication in our experiments arises from a chain of resistor-capacitor
(RC) low-pass filters that is applied to each electrode to reduce
electric field noise from external sources. Due to the high speed
of the fast ion transport waveforms, we must pre-compensate for the
low-pass effects of the filters in order to achieve the desired time-dependence
of the electrode potentials \citep{JPHomeCom}. The filters for each
electrode, from AWG-side to trap-side, are a pair of 200~kHz low-pass
RC filters followed by one 800~kHz low-pass RC filter (see Fig.~\ref{figure:trapfilters}).
The input potential $V_{input}$ to a single stage of the RC filters
in the filter chain are related to the desired output potential $V_{output}$
by

\begin{multline}
V_{input}=V_{output}+R_{i}C_{i}\frac{dV_{output}}{dt}\\
+(V_{output}-V_{prev})\frac{R_{i}}{R_{prev}},\label{eq:Filter}
\end{multline}
where $R_{i}$ and $C_{i}$ are the RC component values of the $i^{{\rm th}}$
layer of the filter chain and $V_{prev}$ and $R_{prev}$ are from
the solution to the previous layer of filtering trap-side. One result
of numerically solving for the effects of the filter system is exemplified
in Fig.~\ref{figure:FilterCompensation}b. The transport waveforms,
including that shown in Fig. \ref{figure:FilterCompensation}b, were
specified by cubic splines that occupied $\sim6$~\% of each channel's
memory.

Despite the filters' time constants on the scale of a few microseconds,
we can still run a waveform to faithfully transport ions in our trap
in 8~$\mu$s. A convenient characterization of the oscillatory motion
is the average occupation of motional quanta $\bar{n}$. In Refs.\citep{10Hanneke,12Gaebler},
motion was initalized to $\bar{n}\sim0.1$ after cooling. However,
ion transport (in this case, separation) was typically on the time
scale of 1~ms, and after separating ions into separate trap zones,
typical oscillatory motion was excited to $\bar{n}\sim5$ which required
re-cooling. Re-cooling, with respect to Fig. \ref{figure:PieChart},
consisted of approximately three milliseconds of Doppler and Raman
sideband cooling. Using the AWG in the experiments in Ref.\citep{12Bowler},
we achieved $\bar{n}\approx0.2$ after single ion transport in 8~$\mu$s
and $\bar{n}\approx2$ after separating two ions into separate trap
zones in 55~$\mu$s. This is not only on a much shorter time scale
but also requires less re-cooling, to be implemented in future quantum
information processing experiments to be able to compare directly
to Fig.~\ref{figure:PieChart}. Different diabatic transport methods
using a waveform generator with a 2.5~MHz update rate were demonstrated
with ions confined in a 1.4~MHz potential well \citep{12Walther}
with similar low-excitation results for transport. However, this update
rate is too slow for the level of control over the accelerations and
velocities that we desire.

\subsection{Waveform branching}

The capability to quickly switch between waveform branches allows
complex sequences of waveforms to be built up from a series of simpler
component waveforms. For example, the experiments in Ref.\citep{12Gaebler}
required two $^{9}$Be$^{+}$ qubit ions to be in a shared trap zone
for two-qubit operations, separated to distinct zones for one-qubit
operations, and recombined; the full sequence required as many as
16 separation and recombination waveforms in a single process. We
used three waveform branches pre-loaded into AWG memory and selected
with TTL inputs during the experiments: (1) the static trap potential,
(2) the separation waveform, and (3) the recombination waveform. The
experimental sequences of these three waveform branches were concatenated
on the fly without the necessity to upload a new set of waveforms
for each sequence, enabled by the AWG branching capability.

\subsection{Shaped Microwave Pulses}

\begin{figure}[t]
\includegraphics{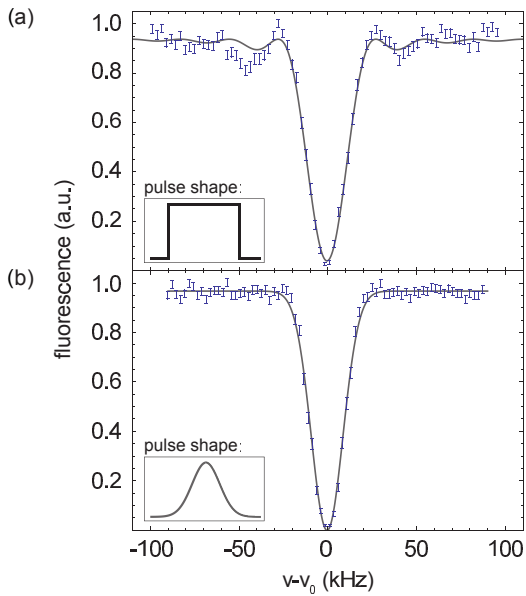}

\caption{Temporal pulse shaping of a microwave signal using the AWG to specify
the profile. The microwave drives the hyperfine transition in $^{25}$Mg$^{+}$.
Shown is the ion fluorescence vs. microwave frequency. Here $\nu_{0}=1.7$~GHz
denotes the frequency splitting of the two hyperfine levels. (a) We
used a rectangular pulse shape, duration 20~$\mu$s, and found the
corresponding sinc$^{2}$ form in frequency space. (b) We used a Gaussian
pulse shape and found the corresponding Gaussian form in frequency
space. The FWHM pulse duration is 45~$\mu$s, with the Gaussian tail
truncated to a total duration of four times the FWHM.}

\label{figure:PulseShaping}
\end{figure}
Another application of the AWG involves the shaping of laser and microwave
pulses for quantum logic. We demonstrate the pulse shaping of a microwave
signal at $\sim1.7$~GHz that drives a hyperfine transition in $^{25}$Mg$^{+}$.
For the experiments in Ref.\citep{11Ospelkaus}, we used the pulse
shaping to prevent \textquotedbl{}ringing\textquotedbl{} of high power
($\sim45$~dBm) pulsed microwave signals during turn on/off. To circumvent
this ringing, we ramped up the power of the signals over 10~$\mu$s.
Using an analog multiplier, one channel from the AWG was used to amplitude
modulate the microwave signal. The waveforms from the AWG can be varied
in duration and amplitude. After this pulse shaping, the signal is
amplified and used to drive the hyperfine transition. In Fig.~\ref{figure:PulseShaping},
we show resonances (in frequency space) of the hyperfine transition
followed by a state-dependent fluorescence measurement for two cases
(here done with $\sim5$~dBm to minimize the impact of ringing):
a rectangular pulse shape (Fig.~\ref{figure:PulseShaping}a), and
a Gaussian pulse shape (Fig.~\ref{figure:PulseShaping}b). We find
that the transition profiles correspond to the Fourier transform of
the pulse shape, a sinc$^{2}$ and a Gaussian, respectively, with
pulse durations chosen for approximately equal full width at half
max (FWHM) for each profile.

\subsection{Shaped Optical Pulses}

Laser beams are commonly used for motional state preparation and generation
of spin-spin entanglement in ion-based quantum information processing.
As the ion-laser interactions are not typically continuous, a computer-controlled
light modulator is used to pulse on and off the laser beams. Acousto-optic
modulators (AOM) can be used as optical switches; actuation is achieved
by a TTL-controlled RF switch. The resulting optical pulse has a rise-time
of $\sim1\,\mu$s, set by the acoustic wave transit time over the
laser beam in the AOM. For some applications, greater control over
the optical pulse amplitude is desired.

A RF mixer is a passive circuit that can be used for high-bandwidth
amplitude modulation of a RF carrier. The amplitude of the modulation
current $\text{{i}}_{{\rm {IF}}}$ applied to the mixer's IF port
varies the RF-port to local oscillator (LO)-port coupling. As the
coupling is inherently nonlinear with $\text{{i}}_{{\rm {IF}}}$,
it is convenient to use closed-loop feedback to obtain a linear response.
In this case the feedback is from a light-sensing photodiode (PD).
A collateral benefit of using a closed loop to linearize the response
is that amplitude noise in the laser beam is attenuated over the loop
bandwidth.

\begin{figure}[t]
\includegraphics{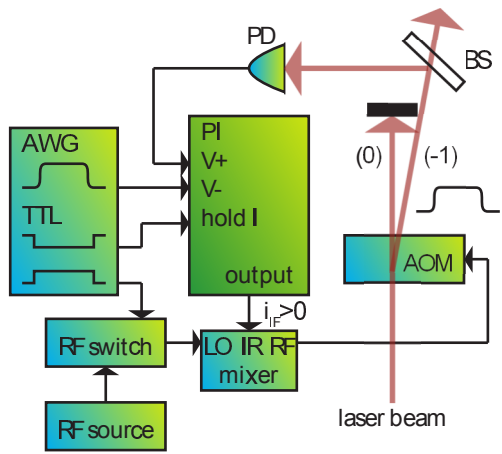}

\caption{Laser beam amplitude modulation circuit using a single DAC channel
of the the AWG, a mixer, and a PI controller in a feedback loop.}

\label{figure:PIServo}
\end{figure}

The current $\text{{i}}_{{\rm {IF}}}$ sent to the mixer's IF port
is determined by an analog proportional-integral (PI) servo with an
integrator track-and-hold (TAH) capability (a schematic of the experimental
set up is shown in Fig.~\ref{figure:PIServo}). The PI is configured
to minimize |$V_{+}-V_{\mbox{-}}$|, where $V_{+}$ is the PD voltage
and $V_{-}$ is an output channel of the AWG. The TAH prevents undesired
activity from the integrator when the laser is ``off'', e.g. due
to non-zero photocurrent from the PD. Clamping $\text{{i}}_{{\rm {IF}}}>0$
prevents bistability of the loop. Laser beam attenuation is controlled
by utilizing the -1 diffracted order of the AOM for the laser (often
in a double-pass configuration). To attenuate the laser beam intensity
beyond that permitted by minimum LO-RF transmission, we employ an
RF switch to extinguish the -1 diffracted order of the AOM.

A near-resonant laser beam is used to laser cool the harmonic motion
of $^{9}$Be$^{+}$ ions ($\omega_{z}/2\pi\sim1$~MHz) with a 10~\%
duty cycle in a Penning Trap with $\sim200$ ions \citep{12Sawyer}.
During Doppler laser cooling, the ion plane is slightly displaced
from the trap center by the cooling light propagating perpendicular
to the plane. When this cooling laser is switched off quickly relative
to $\omega_{z}$ (non-adiabatic), the ion plane oscillates in the
trap with a coherent amplitude corresponding to $\bar{n}$ which is
consistent with a thermal energy of 5-10~mK, well above the Doppler
cooling limit of 0.4~mK. To achieve the Doppler limit for center-of-mass
motion, we ramp the cooling laser intensity adiabatically over $\sim100\,\mu$s
using a $\sin(\omega t)^{2}$ temporal profile, where $\omega/2\pi\sim2.5$~kHz
is the ramp rate.

\section{Summary}

We have designed and constructed an AWG with voltage outputs with
a range of $\pm$10~V and with a 50~MHz update rate. The FPGA provides
versatile control over and storage of waveforms. The output can be
used to control trapping potentials in ion-based experiments, and
several AWG units can be combined to control many trap electrodes.
The AWG can also be used for temporal shaping of laser and microwave
pulses. Branching capabilities controlled by TTL logic allow quick
switching between waveforms with 40~ns latency that can be conditioned
on experimental parameters, a feature useful for error correction
algorithms. We gave some examples of more detailed accounts of the
AWG's capabilities in the applications of high-speed transport of
ions and the shaping of quantum logic pulses.

This work was supported by IARPA, ARO contract no. EAO139840, ONR,
DARPA, and the NIST Quantum Information Program. We thank D. Leibfried,
J. Gaebler and R. J$\text{{ö}}$rdens for helpful comments on the
manuscript. This paper is a contribution of NIST and not subject to
U.S. copyright.

\bibliographystyle{aipnum4-1}

\begin{thebibliography}{20}
\bibitem [{Tra()}]{TradeDeclaration}%
  \BibitemOpen
  \href@noop {} {}\bibinfo {note} {Commercially sourced materials and services
  are identified in this paper in order to specify the designs adequately. Such
  identification is not intended to imply recommendation or endorsement by
  NIST, nor is it intended to imply that the materials or equipment identified
  are necessarily the best available for the purpose}\BibitemShut {NoStop}%
\end{thebibliography}

\end{document}